\pdfoutput=1

\documentclass[data,article,accept,moreauthors,pdftex,10pt,a4paper]{mdpi} 
\firstpage{1} 
\articlenumber{x}
\doinum{10.3390/------}
\pubvolume{xx}
\pubyear{2018}
\copyrightyear{2018}
\history{Received: 16 September 2018; Accepted: 1 October 2018; Published: date}

\Title{Binary Star Database (BDB): New Developments and Applications}

\Author{Oleg Malkov $^{1,}$*\orcidA{}, Aleksey Karchevsky $^{2}$,
Pavel Kaygorodov $^{1}$\orcidC{}, Dana Kovaleva $^{1}$\orcidD{}
and Nikolay Skvortsov $^{3}$\orcidE{}}

\AuthorNames{Oleg Malkov, Aleksey Karchevsky, Pavel Kaygorodov,
Dana Kovaleva and Nikolay Skvortsov}

\address{%
$^{1}$ \quad Institute of Astronomy, Russian Academy of Sciences, Moscow, Russia\\
$^{2}$ \quad Faculty of Physics, Moscow State University, Moscow, Russia\\
$^{3}$ \quad Federal Research Center ``Computer Science and Control'', Russian Academy of Sciences, Moscow, Russia}

\corres{Correspondence: malkov@inasan.ru}

\abstract{Binary star DataBase (BDB) is the database of binary/multiple systems
of various observational types. BDB contains data on physical
and positional parameters of 260,000 components of 120,000 stellar
systems of multiplicity 2 to more than 20, taken from a large variety
of published catalogues and databases. We describe the new features in
organization of the database, integration of new catalogues and
implementation of new possibilities available to users. The development
of the BDB index-catalogue,
Identification List of Binaries (ILB), is discussed.
This star catalogue provides cross-referencing between most popular
catalogues of binary stars.}


\begin{document}

\section{Introduction}

The Binary star DataBase (BDB, bdb.inasan.ru) is the development
of the original database constructed at the Besancon Observatory~\cite{BDBOblak}.
Later BDB was moved from Observatoire de Besan\c{c}on
to the Institute of Astronomy of the
Russian Academy of Sciences~\cite{BDB09,BDB11}.
BDB contains information about binary/multiple stars of all known
observational types taken from several dozens of original catalogs of binary stars.
Collection, parsing and indexing of principal catalogues for all types of binaries
are described in \cite{BDB12,BDB14,BDB-A&C}.
Within the framework of the BDB, cross-identification of the objects
included in it was completed and a three-level identification system
Binary Star Data Base (BSDB)
and a catalogue of object identifiers in double and multiple systems,
Identification list of binaries (ILB),
were created, which made it possible to uniquely identify the systems,
pairs and components, included in BDB. The search for objects in the database
is possible both with the identifier (all major identification systems
are supported) and by parameters. BDB contains data on the physical
and observational parameters of about 260,000 components included in
about 120,000 systems with a multiplicity of 2 to more than 20.
New functionality and advancement of BDB was recently discussed in~\cite{BDB16}.
The aim of this work is to describe in more details recent
improvements that have been made in BDB, discuss its development and
new possible applications of BDB.

BSDB identification system and ILB 
are briefly described in Sections \ref{sec-bsdb} and~\ref{sec-ilb}, respectively.
Observational types of binaries, included in BDB,
are considered in Section~\ref{sec-obstype}.
In particular, there we discuss cataclysmic binaries.
Connection with external databases are described in Section~\ref{sec-link}.
 
\section{Binary and Multiple Star Objects Identification Scheme BSDB}
\label{sec-bsdb}

Support of data on a large number of heterogeneous, but intersecting
observational types of stars (see Section~\ref{sec-obstype})
required conducting of a thorough cross-identification and
construction of a new identification system, BSDB.
The BSDB scheme covers all types of observational data.
Three classes
of objects introduced within the BSDB nomenclature provide correct
links between objects and data, what is especially important for
complex multiple stellar systems.
Within BDB, the entire system, the pair and the component have their
unique BSDB identifier. In particular, BSDB is resistant to cases
when a component is itself resolved into sub-components, and when
a new (distant) component of a system is discovered. In these cases
BSDB allows do not change the identification data and to link
information from sources (catalogues) having different spatial resolution.
Processing of the addition of new components, as well as
details of BSDB construction and application are described
in~\cite{BSDB}.

The principles underlying the BSDB identifier compilation do
satisfy the ``IAU Specifications concerning designations for
astronomical radiation sources outside the solar system''.
The problems typical of binary/multiple-star
designation schemes are basically avoided by the BSDB.

\section{Identification List of Binaries}
\label{sec-ilb}

The Identification List of Binaries (ILB) is a star catalogue
constructed to facilitate cross-referencing between different
catalogues of binary stars. ILB is based on the BSDB identification system
(see Section~\ref{sec-bsdb}), and ILB underlies the BDB search engines.

ILB represents a table of identifications (or running numbers) of
double/multiple systems, pairs and components from the following
catalogues (designation schemes): BSDB, Bayer/Flemsteed,
DM (BD/CD/CPD),
HD,
HIP,
ADS,
WDS,
CCDM,
TDSC,
GCVS,
SBC9,
IGR (and some other X-ray catalogues),
PSR
and Discoverer and number designation (DD).
Description of all these designations can be found in the
Dictionary of Nomenclature of Celestial Objects
({\rm http://cds.u-strasbg.fr/cgi-bin/Dic-Simbad}).
Coordinates
(together with their source code) are also given for each
component.
For each pair, ILB provides information on its observational
type(s).

The content of BDB is not limited to data included in ILB.
Having found the necessary objects according to the data available in ILB,
the user can obtain all the information related to the objects
from the original catalogs (copies of which are integrated into BDB).
However, BDB does not perform any filtering or data evaluation,
providing the user with maximum of available information.

ILB currently contains about 520000 entries: 120000 systems,
140000 pairs and 260000 components.
ILB is regularly updated, improved (bugs and errors are fixes)
and expands as new catalogues are added to BDB.
Identification List of Binaries is described in detail in~\cite{ILB},
and problems of matching and verification of
multiple stellar systems in ILB are discussed
in \cite{Skv17CEUR,Skv17Spri,Skv18Spri}.

\section{Observational Types}
\label{sec-obstype}

\begin{figure}[H]
\centering
\includegraphics[width=10 cm]{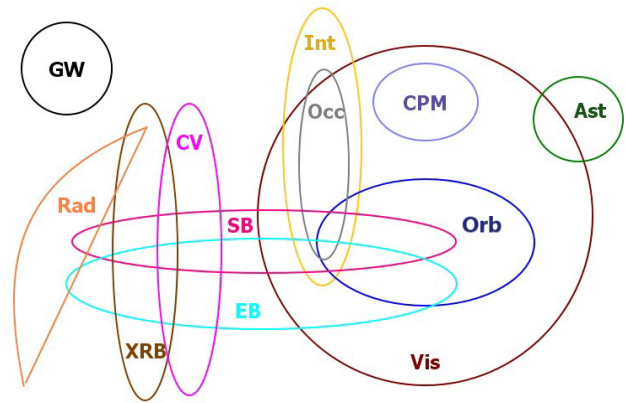}
\caption{Observational types of binaries. See text for details.}
\label{fig:obstypes}
\end{figure}

Observational types of binaries, included in BDB, are illustrated
in Figure~\ref{fig:obstypes}.
Below we give short description of all observational types
(for details see~\cite{BDBObsType} and~\cite{BDBresour}).

\begin{itemize}[leftmargin=*,labelsep=5.8mm]
\item Visual binary ({\bf Vis}) -- a binary star system in which both components
are visible and resolvable in a telescope.
\item Orbital binaries ({\bf Orb}) are visual binaries, demonstrating
sufficiently remarkable orbital motion to calculate orbital parameters.
\item Common proper motion ({\bf CPM}) binaries are pairs of stars
separated by a few arcseconds or arcminutes and having the same proper motions.
The members of each pair presumably formed in close association,
are expected to be coeval and to have the same chemical composition.
\item Astrometric binaries ({\bf Ast}), or stars with invisible companions,
can be deduced from periodic variations in the star's position,
if the position is determined relative to other stars.
The alterations in position, which are superimposed on the proper motion of the star,
are caused by its revolution with the invisible companion
about their common centre.
\item Interferometric binaries ({\bf Int}) present cases
where the components are so close that it can not be observed directly
as a visual binary, but can be resolved using various interferometric techniques.
\item Occultation binaries ({\bf Occ}) are discovered by photoelectric analysis
of lunar (or asteroid) occultation of very close pairs.
\item In spectroscopic binaries ({\bf SB}) the variation of the radial velocities
during a revolution of a binary star can be observed spectroscopically:
the lines in the overlapping spectra of the two components show periodic doubling
or displacement (so called SB2 and SB1, respectively).
\item Eclipsing binaries ({\bf EB}) are binary stars of which
one at times eclipses the other, thus leading to alterations
in the apparent total brightness of the combined stars.
\item X-ray binaries ({\bf XRB}) are pairs of stars producing X-rays,
as the stars are close enough together that material is pulled off
the normal star by the mass of the dense star.
\item Radiopulsars in binary systems ({\bf Rad}) show periodic variations
in the pulsation period caused by orbital motion.
\item Recently detected gravitational-wave signals
produced by the coalescence of two stellar-mass black holes
led to appearance of a new observational type of binaries,
gravitational-wave binaries ({\bf GW}).
\item Cataclysmic variables ({\bf CV}) -- see below.
\end{itemize}

Size of areas in Figure~\ref{fig:obstypes} does not comply with
number of known binaries of the particular observational type.
Overlaps in Figure~\ref{fig:obstypes}
demonstrate that some binaries can be detected by two or more
observational techniques. Thus, an overlap of Vis and  SB areas
contains well known resolved spectroscopic binaries (RSB)
observational type. Note also that currently GW is an isolated
area in our diagram, but later, when
signals from coalescence of two {\it neutron stars} are detected,
it will probably overlap with the XRB area.

\subsection{Cataclysmic Variables}

Cataclysmic variables (CV) are close binaries, undergoing mass transfer
and exhibiting sudden outbursts, generally arising either
from the release of gravitational energy through accretion
or from thermonuclear processes.

Currently we include cataclysmic variables in BDB. Main sources of that
observational type of binaries are
Catalogue of Cataclysmic Binaries, Low-Mass X-Ray Binaries and Related Objects
(Edition 7.24, 31 Dec 2015 -- The Final Edition)~\cite{RK},
Catalog of Cataclysmic Variables (Ver. 2011-2006)~\cite{Downes}, and
General Catalogue of Variable Stars (Ver. GCVS 5)~\cite{GCVS}.
The three catalogues contain 1429, 1830 and 938 CV, respectively, and some
of CV are included in more than one catalogue. Some tens of CV are
included in only one of three catalogues.

We have carried out a cross-identification of CV in the three catalogues
and have made a search of these objects in Simbad.
It should be noted that by no means all catalogued CV are included
in Simbad. Altogether 617 CV are included both in three catalogues
and in Simbad.

Sometimes CV are observed also as binaries of other types.
Among the studied catalogues, only~\cite{RK} contains indication of
eclipsing or/and spectroscopic binarity for the catalogued CV.
Some of objects, catalogued in~\cite{Downes}, are in fact non-CV.

Besides the cataclysmic binaries,~\cite{RK} contains low-mass X-ray binaries
and related objects.
Low-mass X-ray binaries from~\cite{RK} were included in BDB earlier (see~\cite{XRBRad}),
while related objects are being included in BDB at this stage.
There are 619 such objects in~\cite{RK}, and we designate them as pre-CV objects.

We should note that there is a number of other lists of CV published
in literature, however we currently discard them for one or more of
the following reasons: (i) they contain too few objects, (ii) they contain
just {\it candidates} to CV, (iii) they provide too scarce information on
the catalogued objects.

\subsection{Other Observational Types of Binaries}

Some observational types of binaries
are not yet included in BDB and, consequently, are not discussed
in the current paper. Among them are
\begin{itemize}[leftmargin=*,labelsep=5.8mm]
\item chromospherically active binaries,
\item composite spectrum (including symbiotic) binaries,
\item spotted variables (binaries),
\item ellipsoidal variables (binaries),
\item reflecting variables (binaries).
\end{itemize}
All of these types are relatively small
and contain no more than a couple of hundred (two former types)
or just a dozed (the others) of objects.

\section{Connection with External Databases}
\label{sec-link}

Additional information on binary/multiple stars can be obtained from
other databases/catalogues. We have established links to general
purpose databases: Simbad, VizieR, ADS. Corresponding buttons appear
in the BDB resulting page, and one can get information on the queried
objects from the databases.

Besides, we plan to establish links to databases/catalogues of
binaries of particular observational type:
WDS~\cite{WDS} for visual binaries,
SB9~\cite{SB9} for spectroscopic binaries,
CEV~\cite{CEV} for eclipsing binaries, etc.

\vspace{6pt} 

\authorcontributions {The authors made equal contribution to this work; O.M. wrote the paper.}

\funding{This research was partly funded by
the Russian Foundation for Basic Research grant 16-07-01162 and
the Presidium of the Russian Academy of Sciences Program P 28.}
 
\acknowledgments{We are grateful to our reviewers
whose constructive comments greatly helped us to improve the paper.}

\conflictsofinterest{The authors declare no conflict of interest.}


\reftitle{References}





\end{document}